\documentclass[aps,twocolumn,groupedaddress,prl,letter]{revtex4}
\usepackage{epsfig}

\begin{document}
\newcommand{\tb}{ {\bf {t}}}
\bibliographystyle{/usr/share/texmf/bibtex/bst/revtex4/apsrev}

\title{Scintillation of liquid neon from electronic and nuclear recoils}

\author{J. A. Nikkel}
\email{james.nikkel@yale.edu}
\author{R. Hasty}
\author{W. H. Lippincott}
\author{D. N. McKinsey}
\affiliation{Yale University, New Haven, CT}

\date{\today}

\begin{abstract}

We have measured the time dependence of scintillation light from
electronic and nuclear recoils in liquid neon, finding a slow time constant of
$15.4 \pm 0.2$~$\mu$s. 
Pulse shape discrimination is investigated as a means of identifying event type in
liquid neon.   Finally, the nuclear recoil scintillation efficiency is measured to be
$0.26~\pm~0.03$ for 387~keV nuclear recoils.

\end{abstract}

\maketitle

\section{Introduction}

Liquid neon has been suggested as a low-background scintillation medium for the
CLEAN (Cryogenic Low Energy Astrophysics with Noble gases) 
experiment, which will have simultaneous sensitivity to both low energy
solar neutrinos and WIMPs~\cite{McKinsey:2000, McKinsey:2005b,
Boulay:2005}. The CLEAN detector will use approximately 100~tonnes of liquid
neon surrounded by photomultiplier tubes (PMTs) in a spherical geometry.  Scintillation
light resulting from both electronic and nuclear recoils
will be detected by these PMTs. Liquid neon is a good candidate for low background
experiments because it has no long lived radioactive isotopes, is easily
purified, and is dense enough to allow for self-shielding, effectively
eliminating backgrounds in the center of the detector. The solar neutrino
signal will consist of electronic recoils, while a WIMP signal will consist
of nuclear recoils.  The physics goals of CLEAN require a method of
discriminating between these two types of events.     

It was pointed out in Ref.~\cite{McKinsey:2005b} and \cite{Boulay:2005} that
electronic and nuclear recoil events in liquid neon might be distinguished
through the use of pulse shape discrimination (PSD).
Interactions of ionizing radiation with liquid noble gases produce excimers that can
exist in either a singlet or triplet state.  The lifetime of the triplet
state in liquid neon has been previously measured to be
3.9~$\mu$s~\cite{Michniak:2002}.  The singlet lifetime is expected to be much
shorter, on the order of a few nanoseconds, based on measurements of other
liquid noble gases~\cite{Hitachi:1983}.

The slow scintillation light emitted by triplet molecules is suppressed in 
intensity by destructive triplet interactions, primarily
Penning ionization and electron-triplet spin exchange; these reactions are
stronger for higher excitation densities such as those produced by nuclear
recoils.  For a given event, measurement of the fraction
of the total scintillation signal that is in the fast component can be
used to determine which type of event has occurred.

PSD has also been studied in detail for liquid xenon based
experiments~\cite{Davies:1994, Akimov:2002} and can be used to
suppress gamma ray backgrounds.  PSD in liquid helium has also been studied in order
to separate electronic recoil events from $^{3}$He(n,p)$^{3}$H events in
the search for the neutron electric dipole moment~\cite{Golub:1994}.
In an extension of the CLEAN concept, PSD has been also suggested to counter
$^{39}$Ar background in single phase liquid argon dark matter
detectors~\cite{Boulay:2006}. Previous measurements of liquid and solid
neon~\cite{Michniak:2002} have compared beta induced scintillation to 5 MeV
alpha induced scintillation, but no studies exist of PSD for the nuclear
recoils characteristic of a WIMP signal. 

Because the WIMP-nucleon scattering rate is expected to decay
exponentially with nuclear recoil energy, the sensitivity of
experiments to WIMPs depends crucially on energy
threshold, which in turn depends both on the scintillation yield for nuclear
recoils and the light detection efficiency of the apparatus. The scintillation
yield from nuclear recoils relative to that from electronic recoils has been
measured recently by several groups for liquid xenon
~\cite{Arneodo:2000a,Akimov:2002,Aprile:2005,Chepel:2006}.  In contrast, the
scintillation yield for nuclear recoils in liquid neon has not been measured.  

In this paper we describe measurements of the pulse shape discrimination
efficiency, nuclear recoil scintillation efficiency, and time dependence of
liquid neon scintillation.

\section{Experimental Details}
\label{sec:exp_det}

\begin{figure}[ht]
  \centerline{
    \hbox{\psfig{figure=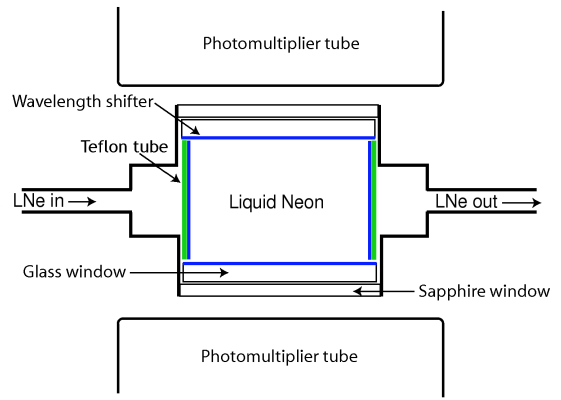,width=7.5cm,%
                clip=}}}
          \caption[Scintillation cell]
                  {Schematic representation of the scintillation cell.}
          \label{fig:cell}
\end{figure}

The data presented here were obtained using a 200~ml
scintillation vessel viewed by two 75~mm diameter photomultiplier
tubes (PMTs)~\footnote{Electrontubes model D746,
www.electrontubes.com}.  A schematic of the setup can be seen in
Fig.~\ref{fig:cell}.
 
The scintillation vessel is a stainless steel body attached to two titanium
flanges into which 50~mm diameter sapphire windows are bonded with Stycast~2850
epoxy.  Titanium and sapphire were chosen due to their 
similar thermal expansion coefficients, allowing a leak-tight seal when filled
with liquid neon. Because liquid neon scintillates in the ultraviolet 
($\approx~77 $~nm)~\cite{Packard:1970}, we place inside the vessel secondary windows
coated with a thin film of
tetraphenyl butadiene (TPB~\cite{McKinsey:1997}).  
The TPB shifts the wavelength of the ultraviolet light to approximately 440~nm so that
it may pass through the 
windows and be detected by the PMTs. The top window is coated with
0.29~mg/cm$^2$ of TPB and the bottom window with 0.18~mg/cm$^2$. The inside wall of the vessel is lined
with a PTFE sheet that is coated on the inner surface with
0.16~mg/cm$^2$ of TPB. 

The scintillation vessel is housed inside a vacuum dewar for thermal insulation, and liquid neon is
introduced though a tube on the side of the scintillation vessel.
The neon is liquefied from purified gas in a copper can mounted to the end of
a pulse-tube refrigerator~\footnote{Cryomech model PT805, www.cryomech.com}.   
All components that come into contact with the gas or liquid are baked to at
least 60~C for 12 hours, and the ultra-high purity neon gas (99.999\%) is passed
through a heated 
gas purification getter~\footnote{Omni Nupure III, www.nupure.com} before
it is liquefied.

A stainless steel tube leading out of the scintillation vessel leads to a
charcoal trap connected in line.  Purification of neon with
charcoal using this method has been previously discussed in Ref.~\cite{Harrison:2006}.
After the cell is filled with liquid, a small heater below the trap is energized, forcing
vaporized neon through the charcoal and back into the liquefaction can.  By
monitoring the heat load on the refrigerator we calculate that a
continuous flow rate of approximately 0.5 gas litres per minute 
was obtained over the course of the experiment. 

\begin{figure}[ht]
  \centerline{
    \hbox{\psfig{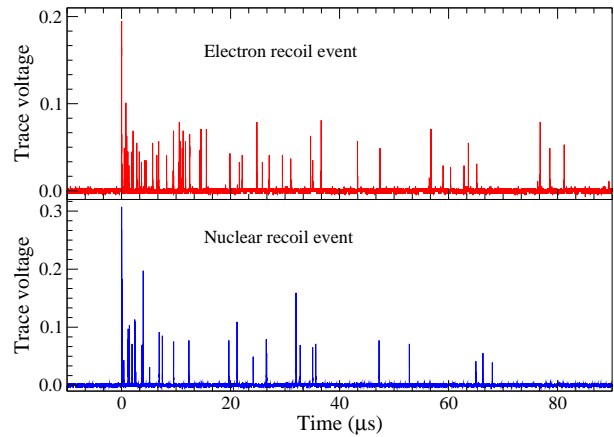}}}
          \caption[Sample scintillation trace]
                  {Examples of two events that are identified as 
                    electronic and nuclear recoils using pulse shape
                    discrimination.  The discrimination method used is
                    described in the Experimental Results section.}
          \label{fig:trace_ex}
\end{figure}

Two sample oscilloscope traces from 
scintillation events in liquid neon can be seen in Fig.~\ref{fig:trace_ex}.
The signal from each PMT is divided three ways with a linear fan-out. 
One copy goes to the triggering electronics and two copies go
directly to an oscilloscope where they are digitized at two different
gain settings.  We use this technique to increase the effective
dynamic range of the 8-bit oscilloscope digitization.  The triggering
electronics require a coincidence of at least one half of a
photoelectron (pe) within 100~ns in each PMT. 

All data analysis is performed in software.
The first analysis step is to take collections of runs, typically two hours
long, and find the value of the single photoelectron signal.  For all types
of excitations in the neon, there is a significant triplet component, and
since this is spread out over many $\mu$s, a large number of single
photo-electron events can be analyzed to obtain the gain of the
PMTs.

\subsection{Sodium-22 Calibration}
\label{sec:na22}

\begin{figure}[h]
  \centerline{
    \hbox{\psfig{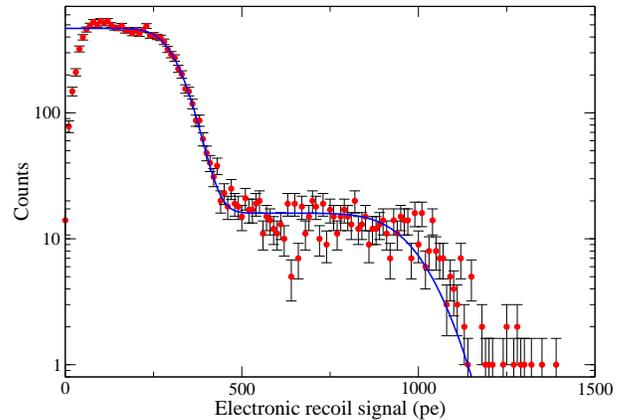}}}
          \caption[Plot of $^{22}$Na data]
                  {Plot of $^{22}$Na calibration data with fit to
                    Eq.~(\ref{eq:na22_fit}), indicating a
                    detector signal yield of $0.93~\pm~0.07$~photoelectrons per
                    keV.}
          \label{fig:na22data}
\end{figure}

We use a 10~$\mu$Ci sealed $^{22}$Na source
to determine the absolute signal yield for electronic recoils. 
This source is placed on the outside of the vacuum enclosure to
provide a source of high energy gamma rays.  Because the gammas from this source
are produced in the  $e^+ + e^- \rightarrow \gamma + \gamma $ reaction, we can
significantly reduce backgrounds due to other radioactive decays, as well as
cosmic rays, by requiring a coincidence between the scintillation vessel and
an external organic scintillation detector. The majority of the 511~keV gamma
rays that interact in the neon Compton 
scatter, providing a wide range of energy depositions over the 
region of experimental interest.  We also use these `tagged' events for measuring the
electronic recoil time dependence.

Data are analyzed from the two PMTs with the only
requirement that the integrated signal from one PMT be within a factor of
four of the other.  We fit the calibration data to a function that is
comprised of two Gaussian-broadened step-functions,

\begin{eqnarray}
  N &=& A\left[1-\mathrm{erf}\left({\frac{E-E_{511}}{\sigma_1\sqrt{2}}}\right)\right] \nonumber \\
  &+& B\left[1-\mathrm{erf}\left({\frac{E-E_{1274}}{\sigma_2\sqrt{2}}}\right)\right],
  \label{eq:na22_fit}
\end{eqnarray}
where $E=x/Y$ is the energy, $Y$ is the yield, $E_{511}$ is the value of
the 511~keV Compton edge (341~keV) and $E_{1274}$ is the value of
the 1274~keV Compton edge (1062~keV).  
$\sigma_1$ and $\sigma_2 (=\sigma_1 \sqrt{E_{511}/E_{1274}})$ are
the widths, and $A$ and $B$ are the relative sizes of the steps.
The calibration data are shown in Fig.~\ref{fig:na22data} along with a four
parameter fit for $\sigma_1$, $A$, $B$, and $Y$.  From this fit, we have
determined that the signal yield for electronic recoils in this detector is
$0.93~\pm~0.07$~pe/keV at 511~keV.  

We measured the stability of the system by taking individual $^{22}$Na runs
over the course of four days.  We found that the signal yield remained
constant to within  $7\%$.

\subsection{Nuclear recoils}
\label{sec:nuc}

\begin{figure}[h]
  \centerline{
    \hbox{\psfig{figure=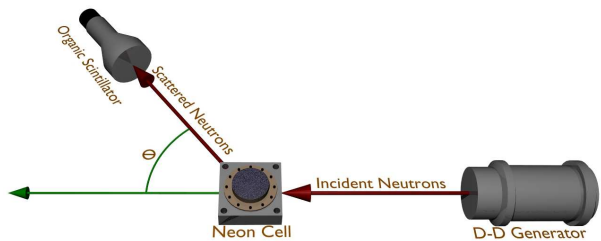,width=8cm,%
                clip=}}}
          \caption[Neutron scattering schematic]
                  {Schematic of the neutron scattering setup.}
          \label{fig:nuc_scatt}
\end{figure}

To investigate the detector response to nuclear recoils, we use a portable
deuterium-deuterium neutron generator~\footnote{Themo Electron Model MP320,
www.thermo.com} as a source and an organic scintillator as a secondary
detector. The experimental setup can be seen schematically in
Fig.~\ref{fig:nuc_scatt}.  The nuclear recoil energy, $E_{rec}$,
may be determined from the angle of the scattered neutron through
simple kinematics, 

\begin{eqnarray}
  \setlength{\arraycolsep}{-0.2cm}
  E_{rec} &=& \frac{2 E_{in}}{(1+M)^2} \left[ 1 +  M - \cos^2(\theta) \right. \nonumber \\
    &-& \left. \cos(\theta) \sqrt{M^2+\cos^2(\theta)-1}  \right],
    \label{eq:nuc_recoil}
\end{eqnarray}
where $E_{in}$ is the incident neutron energy (2.8~MeV),
$M$ is the atomic mass (20.18 for neon), and $\theta$ is the scattering
angle of the outgoing neutron.

\section{Experimental Results}

\subsection{Pulse shape discrimination}
\label{sec:PSD}

To discriminate between electronic recoil and nuclear recoil events,  
we calculate $f_{prompt}$, the fraction of the signal that arrives in the
first 100~ns.   

\begin{figure}[ht]
  \centerline{
    \hbox{\psfig{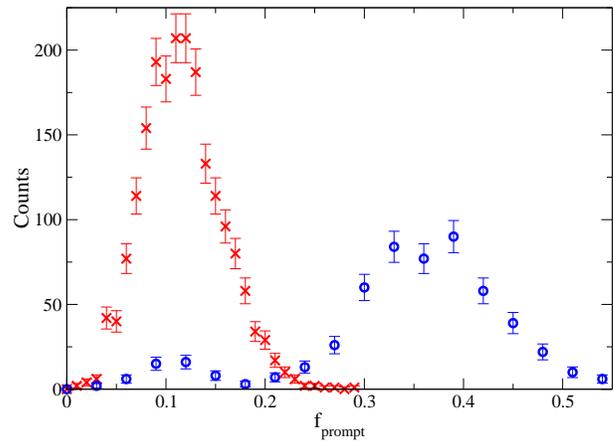}}}
          \caption[$f_{prompt}$ plot]
                  {Plot of the fraction of light in the prompt component
                  ($f_{prompt}$) for electronic recoils (Xs) and
                  electronic plus nuclear recoils (circles). 
                  Events are selected to be between 80 and 120 photoelectrons
                  for this plot.}
          \label{fig:f_prompt}
\end{figure}

Figure~\ref{fig:f_prompt} is a histogram of $f_{prompt}$ for 1434 events
identified as  electronic recoils and 573 events identified as nuclear
recoils. All events are selected to be between 80 and 120 photoelectrons.
As one can see, there is a natural division at about $f_{prompt}=0.25$.  Using a
50\% nuclear recoil acceptance ($f_{prompt} \geq 0.35$) we conclude that the
discrimination efficiency exceeds one part in 1434.

There is a small population of events in the nuclear recoil set that have a
small $f_{prompt}$.  The neutron generator produces a large number of gamma rays, and some
of those are misidentified by the time of flight cuts.

\subsection{Time dependence}
\label{sec:tau}

\begin{figure}[h]
  \centerline{
    \hbox{\psfig{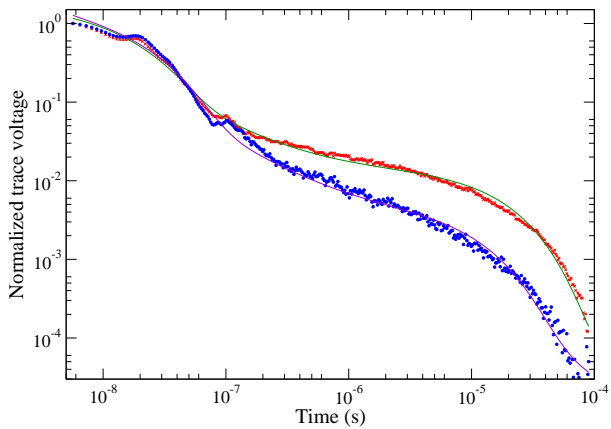}}}
          \caption[Time dependence]
                  {Plot of the probability density function as a function of
                    time for both electronic recoils (top curve) and nuclear
                    recoils (bottom curve).  Solid lines
                    are fits described in the text.}
          \label{fig:taus}
\end{figure}

We estimate the temporal probability density functions for nuclear and
electronic recoil events in liquid neon based on the data from events that
yield between 80 and 120 photoelectrons.  Because the d-d neutron generator
emits significant numbers of gamma rays, we require that the fraction of light
within the first 100~ns ($f_{prompt}$) be greater than 0.18 before it is considered
to be a nuclear recoil.  There is a clear separation in the neutron and gamma
populations in Fig.~\ref{fig:f_prompt} that justifies this cut.
Additionally, a 100~ns window in the time of flight spectrum is 
allowed in order to reduce the number of gamma ray events in the region of
interest.  No such cuts are used for the tagged $^{22}$Na electronic recoil events. 

For each event class, the trigger time is defined as the time that the voltage rises above 20\% 
of its maximum value.  These aligned pulses are averaged and then normalized in
order to more easily compare them.
Both gamma and neutron data are simultaneously fit using a least squares method to the
following function, 

\begin{eqnarray}
  p_m &=& A  e^{-(t-t_0)/\tau_1} \nonumber \\ 
  &+& \frac{B}{t-t_0} \nonumber  \\
  &+& C e^{-(t-t_0)/\tau_2},
  \label{eq:tau_fit}
\end{eqnarray}
while allowing $A$, $B$, and $C$ to vary between the two different types
of excitations. 

This function was constructed as an extension of the two exponential fit
that one needs as a minimal description of the singlet and triplet decay constants for the
neon molecules.   The quality of fit to the neon data is statistically better with this hybrid
function than it is with either a two or three term exponential fit function.
The motivation for adding the $1/t$ term is based on
observations of radiative decay in liquid helium
scintillations~\cite{McKinsey:2003} where
the non-exponential component is explained as a two-body
reaction effect.  This is a first order approximation to the scintillation
time dependence for triplet-triplet interactions calculated by King~{\it et al.}~\cite{King:1966}. 
The resulting fit parameters are listed in Table~I.
We find the slow time constant to be significantly longer than that measured by Michniak
\textit{et al.}~\cite{Michniak:2002} ($15.4 \pm 0.2$~$\mu$s vs. $3.9 \pm 0.5$~$\mu$s).
We attribute this discrepancy to our continuous purification procedure which
reduces impurities that may quench the triplet molecules.  

The singlet decay time can not be directly calculated from $\tau_1$ as the
speed of the prompt pulse is also determined by the PMT timing and TPB
wavelength shifter lifetime.

\begin{table}[h]
  \begin{tabular}{|c||c|c|}
    \hline
    Parameter & Electronic recoils & Nuclear recoils                   \\
    \hline
    $A$               & $0.86 \pm 0.16$     & $1.33 \pm 0.13$        \\
    $B$ (ns)          & $5.0 \pm 0.2$        & $3.78 \pm 0.08$       \\
    $C$               & $0.0137 \pm 0.0004$    &  $0.00230 \pm 0.00002$  \\
    $t_0$ (ns)        & \multicolumn{2}{c|}{$-4.8 \pm 0.9$}          \\
    $\tau_1$ (ns)     & \multicolumn{2}{c|}{$18.6 \pm 0.2$}          \\
    $\tau_2$ ($\mu$s) & \multicolumn{2}{c|}{$15.4 \pm 0.2$}          \\
    \hline   
  \end{tabular}
  
  \label{tab:tau_fit}
  \caption[Decay fit parameters]
  {Fit parameters to Eq.~(\ref{eq:tau_fit}) for both
    electronic and nuclear recoil events between 80 and 120
    photoelectrons.}
\end{table}

\subsection{Nuclear recoil scintillation efficiency}
\label{sec:nuc_eff}

\begin{figure}[h]
  \centerline{
    \hbox{\psfig{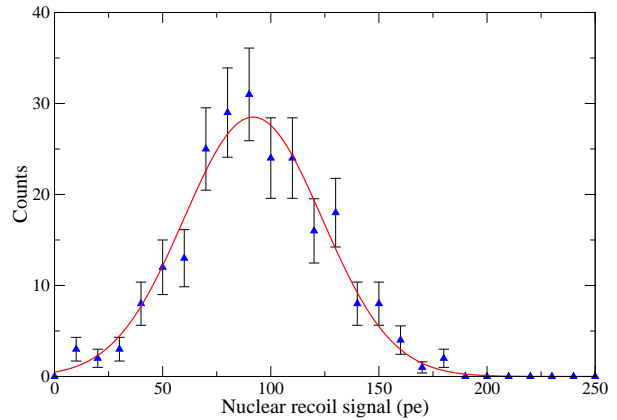}}}
          \caption[Nuclear recoil efficiency]
                  {Plot of the light production for 387~keV~nuclear
                    recoils with Gaussian fit.}
          \label{fig:nr_eff}
\end{figure}

The relative nuclear recoil scintillation efficiency is the ratio of light
produced by nuclear recoils compared to that produced by electronic
recoils of the same energy.  For simplicity we assume that the electronic
recoil scintillation efficiency is constant over our region of interest.

We choose a scattering angle of $120^\circ~\pm~3^\circ$ which corresponds to a
nuclear recoil energy deposition of $387~\pm~11$~keV.  We exclude events from
the raw data set that lie outside of a 60~ns time of flight window.  The
resulting data are plotted in Fig.~\ref{fig:nr_eff}.

The Gaussian best fit parameters yield a peak at $93~\pm~2$ photoelectrons
and a width of $33~\pm~3$ photoelectrons.  Therefore, the nuclear recoil
signal yield is $0.24~\pm~0.01$ photoelectrons per keV.  Using
$0.93~\pm~0.07$~pe/keV for 511~keV electronic recoils, the relative nuclear
recoil scintillation efficiency is $0.26~\pm~0.03$ at $387~\pm~11$~keV.

\section{Conclusion}

We have measured the time dependence of electronic and nuclear recoil
scintillation events in liquid neon.  We find that the slow time constant is
$15.4 \pm 0.2$~$\mu$s and find evidence for a  $1/t$ term through least squares fitting.  
We have also found that in our apparatus the signal yield for 511~keV
electronic recoils is $0.93~\pm~0.07$~pe/keV and for 387~keV nuclear recoils
is $0.24~\pm~0.01$~pe/keV.  This implies a relative scintillation
efficiency of $0.26~\pm~0.03$ at this energy.  For a signal yield between 80
and 120 photoelectrons we find the pulse shape discrimination between nuclear
and electronic recoils to be better than 1 part in 1434 with 50\% nuclear
recoil acceptance.

\bibliography{main_bib}

\end{document}